\documentclass[aps,prl,amsmath,amssymb,10pt,superscriptaddress,twocolumn,color,epsfig,graphicx,bm,floatfix]{revtex4-1}

\usepackage{color}
\usepackage{graphicx}
\usepackage{graphics}
\usepackage{subfigure}
\usepackage{epsfig}
\usepackage{amsmath}
\usepackage{array}
\usepackage{amssymb}
\usepackage{graphicx}      
\usepackage{dcolumn}       
\usepackage{bm}            
\begin{document}
\title{High pressure structures of disilane and their superconducting properties}

\author{Maximilian Amsler}
\affiliation{These authors contributed equally to this work.}
\affiliation{Department of Physics, Universit\"{a}t Basel,
Klingelbergstr. 82, 4056 Basel, Switzerland}

\author{Jos\'e A. Flores-Livas}
\affiliation{These authors contributed equally to this work.}
\affiliation{Universit\'e de Lyon, F-69000 Lyon, France and 
LPMCN, CNRS, UMR 5586, Universit\'e Lyon 1, F-69622 Villeurbanne, France}

\author{Thomas J. Lenosky}
\affiliation{C8 Medisensors, Los Gatos, California 95032, USA}
\author{Lauri Lehtovaara}
\affiliation{Universit\'e de Lyon, F-69000 Lyon, France and 
LPMCN, CNRS, UMR 5586, Universit\'e Lyon 1, F-69622 Villeurbanne, France}
\author{Silvana Botti}
\affiliation{Universit\'e de Lyon, F-69000 Lyon, France and 
LPMCN, CNRS, UMR 5586, Universit\'e Lyon 1, F-69622 Villeurbanne, France}
\author{Miguel A.L. Marques}
\email{miguel.marques@univ-lyon1.fr}
\affiliation{Universit\'e de Lyon, F-69000 Lyon, France and 
LPMCN, CNRS, UMR 5586, Universit\'e Lyon 1, F-69622 Villeurbanne, France}
\author{Stefan Goedecker}
\email{stefan.goedecker@unibas.ch}
\affiliation{Department of Physics, Universit\"{a}t Basel,
Klingelbergstr. 82, 4056 Basel, Switzerland}

\date{\today}

\begin{abstract}
A systematic \textit{ab initio} search for low enthalpy phases of
disilane (Si$_2$H$_6$) at high pressures was performed based on the
minima hopping method. We found a novel metallic phase of disilane with
$Cmcm$ symmetry, which is enthalpically more favorable than the
recently proposed structures of disilane up to 280\,GPa, but revealing  
compositional instability below 190~GPa. The $Cmcm$
phase has a moderate electron-phonon coupling yielding a
superconducting transition temperature $T_c$ of around 20\,K at
100\,GPa, decreasing to 13\,K at 220\,GPa. These values are an order
of magnitude smaller than previously predicted $T_c$ for disilane, and
cast strong doubts on the possibility of high-$T_c$ superconductivity
in these systems as well as in other hydrogen-rich compounds under moderate
pressure. 
\end{abstract}

\pacs{somepacs}

\maketitle

Superconductivity in elemental hydrogen was predicted by
Ashcroft~\cite{ashcroft_metallic_1968} already in 1968. More recently,
and with the use of novel theoretical techniques~\cite{Luders,Marques}, 
the calculated $T_c$ was estimated to be as high as 240\,K at pressures of around
450\,GPa~\cite{cudazzo_ab_2008}. However, the synthesis of metallic
hydrogen has been found to be experimentally challenging, and even at
extremely high pressures (below 342\,GPa) metallization has not yet been
observed~\cite{narayana_solid_1998,LeToullec2002}. This is in
agreement with theoretical calculations, that predict the metallic
transition above 400\,GPa --- a pressure beyond the reach of
current experimental capabilities.

To circumvent this problem, it was recently suggested that
metallization pressures could be achieved in hydrogen rich materials
where the hydrogen is chemically
``pre-compressed''~\cite{ashcroft_hydrogen_2004}.  Several
investigations of such compounds have appeared in the literature,
primarily focusing on group IV-hydrides. Calculations on phases of
highly compressed
silane~\cite{pickard_high-pressure_2006,feng_structures_2006,yao_superconductivity_2007,chen_superconducting_2008,kim_crystal_2008,martinez-canales_novel_2009},
germane~\cite{gao_superconducting_2008} and
stannane~\cite{tse_novel_2007,gao_high-pressure_2010} have shown the
possibility of metallic phases with high $T_c$ at moderate
pressures. From the experimental point of view, silane SiH$_4$ has
been reported to crystallize and attain metallicity above
50--60\,GPa~\cite{eremets_superconductivity_2008,chen_pressure-induced_2008}
with a superconducting behavior. However, more recent studies ascribe
the observed metallicity to the formation of metal
hydrides~\cite{degtyareva_formation_2009}, and metallization of silane
was found not to occur at least below
130~GPa~\cite{hanfland_high-pressure_2011}.

Another hydrogen rich compound of the same family is disilane
Si$_2$H$_6$.  This compound has attracted attention as a hydrogen rich
material due to its experimental availability. Moreover, in a recent theoretical
study, Jin~\textit{et al.}~\cite{jin_superconducting_2010} performed 
random searches in order to find stable structures
of disilane. They reported three different structures covering
a pressure range from 50 to 400\,GPa. Crystallization of disilane into a metallic phase 
with a $P\textendash 1$ lattice was predicted to occur at 135\,GPa. 
The $T_c$ of this phase was predicted to be
64.6\,K at 175\,GPa, and 80.1\,K at 200\,GPa. Beyond 175\,GPa, the
lowest enthalpy phase becomes a cubic $Pm \textendash 3m$ structure that
reaches the remarkable superconducting transition temperature of $T_c
= 139$\,K at 275\,GPa, a $T_c$ much higher than any other
predicted transition temperature of group IV-hydrides. Unfortunately,
these results have not been experimentally confirmed.

\begin{figure}[t]            
\subfigure[]{\includegraphics[width=0.48\columnwidth,angle=0]{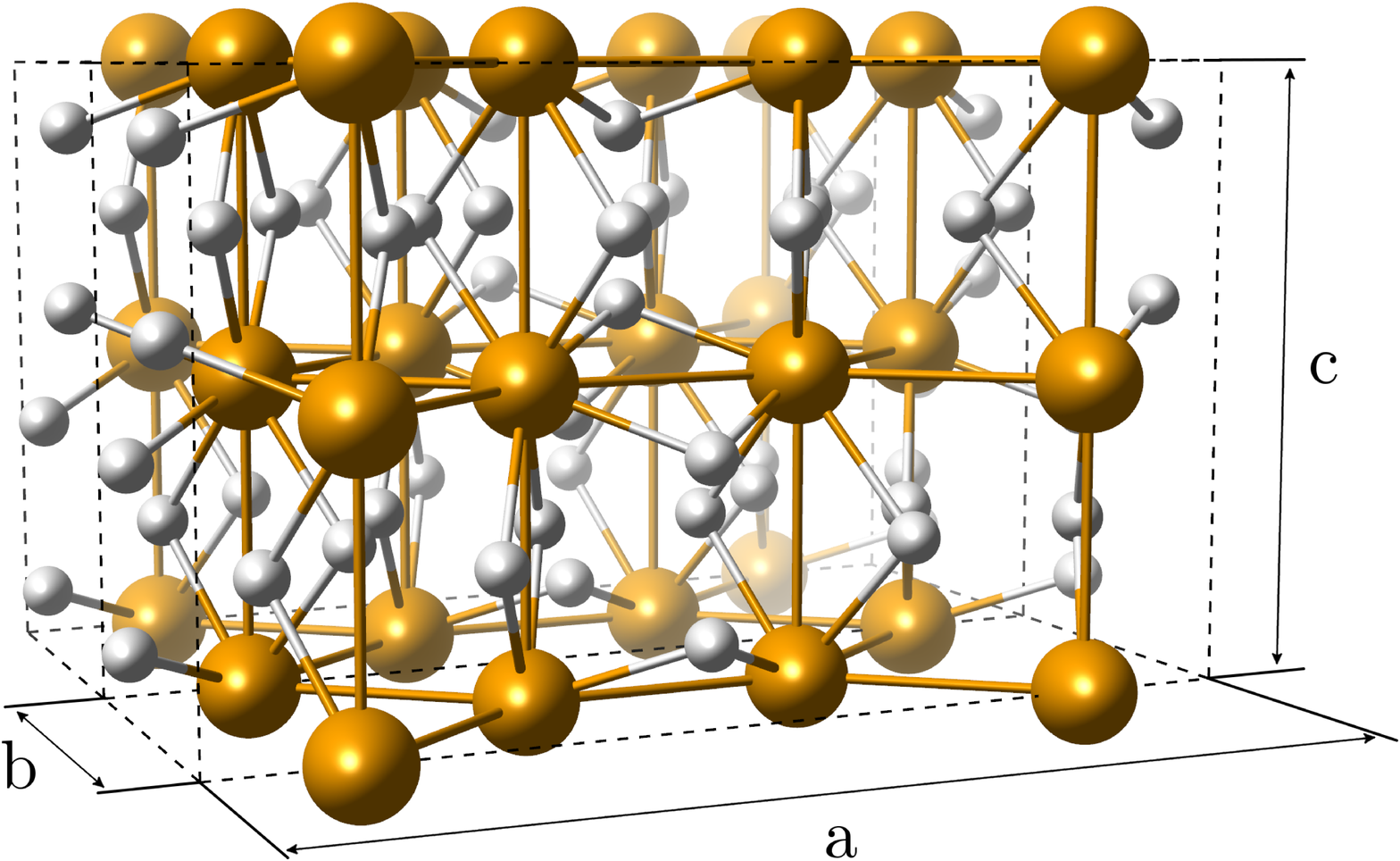}} 
\subfigure[]{\includegraphics[width=0.48\columnwidth,angle=0]{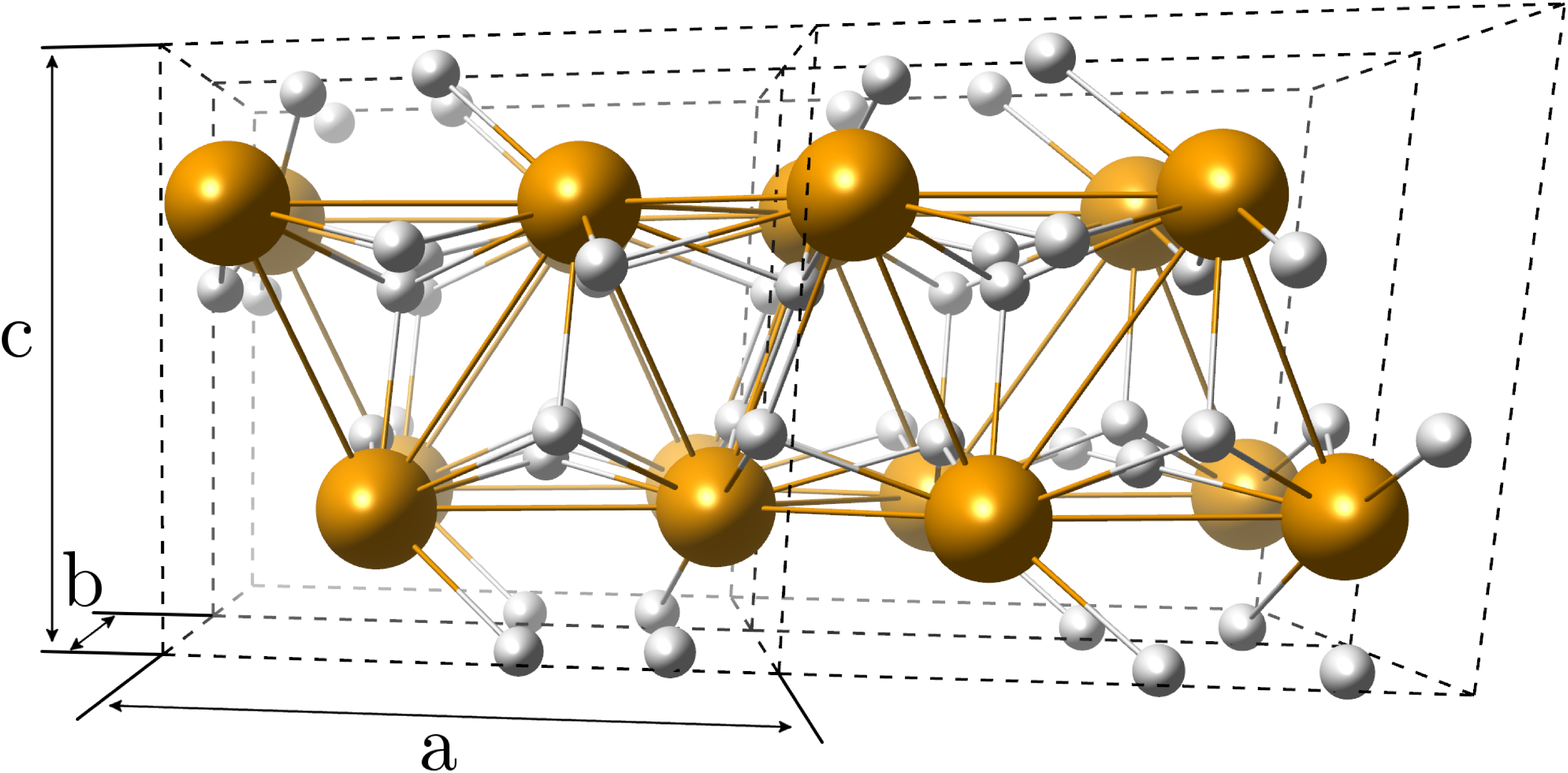}}  
\subfigure[]{\includegraphics[width=0.49\columnwidth,angle=0]{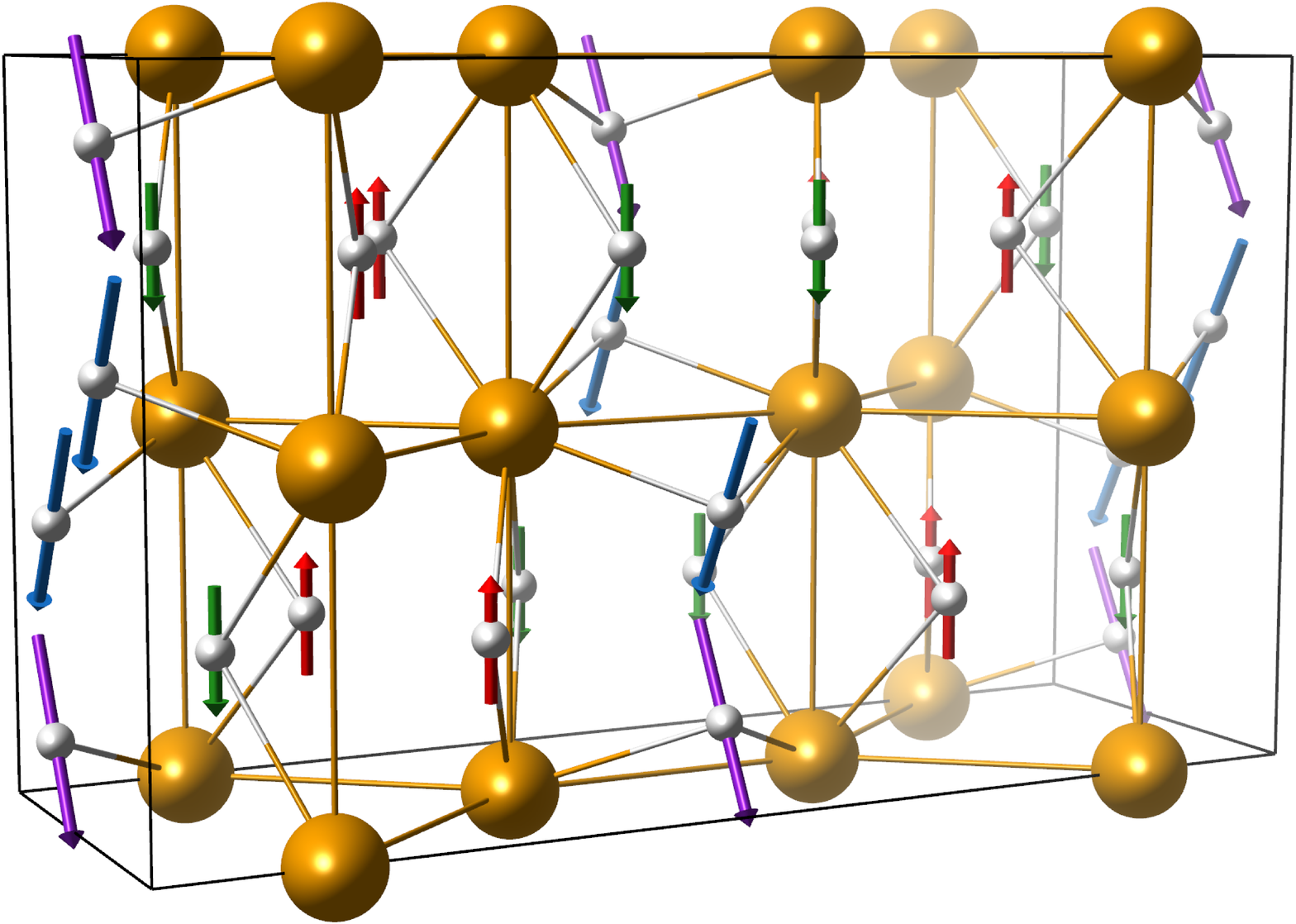}}  
\caption{(Color online) The crystal structures of (a) the $Cmcm$ phase
  at 200\,GPa, and (b) the $P\textendash 1$ phase at 300~GPa. The eigendisplacements which
  lead from the $Cmcm$ structure to the $Cmc2_1$ structure are visualized by
  arrows in panel (c).}
\label{fig:Structure}
\end{figure}

In this letter, we report on our investigations of the disilane system under pressure by
using the recently developed minima hopping method
(MHM)~\cite{goedecker_2004,amsler_2010} for the prediction of low
enthalpy structures.  This method has been successfully used for
global geometry optimization in a large variety of
applications~\cite{hellmann_2007,roy_2009,bao_2009,willand_2010,amsler_crystal_2011}.
Given only the chemical composition of a system, the MHM aims at
finding the global minimum on the enthalpy surface, while gradually
exploring low lying structures. Moves on the enthalpy surface are
performed using variable cell shape molecular dynamics with initial
velocities approximately chosen along soft mode directions. The
relaxations to local minima are performed by the fast inertia
relaxation engine (FIRE)~\cite{bitzek_2006} taking into account both
atomic and cell degrees of freedom.

We performed simulations for cells containing 1, 2, and 3 formula
units of disilane Si$_2$H$_6$ under several different pressures
between 40--400\,GPa. The initial sampling of the enthalpy surface was
carried out employing the MHM together with Lenosky's tight-binding
scheme~\cite{lenosky_highly_1997}, extended to include hydrogen. The most promising
candidate structures found during the initial sampling were further
studied~\cite{gonze_brief_2005,gonze_abinit_2009} at the density functional theory (DFT) level
using the Perdew-Burke-Erzernhof (PBE) exchange-correlation
functional~\cite{PBE96} and norm-conserving
HGH-pseudopotentials~\cite{Pseudopotential}. The plane-wave cut-off energy was
set to 1400\,eV, and carefully converged Monkhorst-Pack k-point
meshes~\cite{M_And_P_Kmesh} with grid spacing denser than
2$\pi\times0.025$\,\AA\ were used. Finally, in order to confirm that
the tight-binding scheme was able to sufficiently sample the enthalpy
surfaces, we performed MHM simulations for selected pressures of
100\,GPa, 200\,GPa, 280\,GPa, and 320\,GPa at the DFT level.

\begin{figure}[t]            
  \includegraphics[width=1.0\columnwidth]{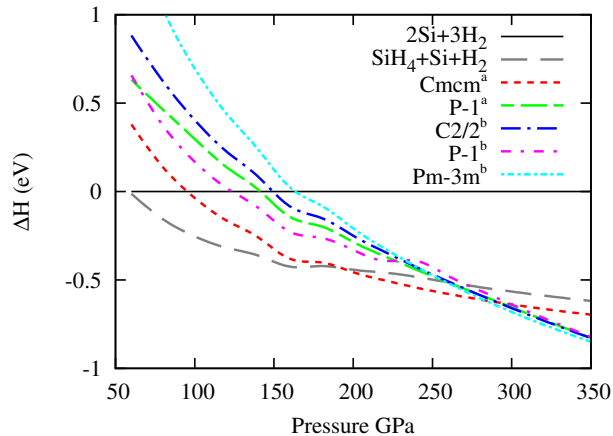}  
  \caption{(Color online) Enthalpy per formula unit of disilane as a
    function of pressure with respect to elements in their solid form
    $2{\rm Si}(s) + 3{\rm H}_2(s)$. The decomposition enthalpies were
    computed from the predicted structures of
    hydrogen~\cite{pickard_structure_2007} and high pressure phases of
    silicon~\cite{olijnyk_structural_1984,duclos_hcp_1987}.  The
    disilane structures with superscripts $^{\textrm{\footnotesize
        a}}$ and $^{\textrm{\footnotesize b}}$ are from this work and
    from Ref.~\onlinecite{jin_superconducting_2010}, respectively.}
  \label{fig:Enthalpy}
\end{figure}

In Fig.~\ref{fig:Enthalpy} the enthalpy of the different phases found
in our MHM simulations are shown with respect to decomposition towards
elemental silicon and hydrogen.  At pressures above 280\,GPa, the
$Pm \textendash 3m$ phase is favored, competing with several other
structures reported by Jin~\textit{et
  al.}~\cite{jin_superconducting_2010}. In addition to the structures 
reported in Ref.~\cite{jin_superconducting_2010} our simulations revealed 
another low-lying phase with $P\textendash 1$ symmetry
(Fig.~\ref{fig:Structure}b). However, all these structures lie in a
very small energy range which is within our numerical precision.
Furthermore, as seen in Fig.~\ref{fig:Enthalpy}, crystalline disilane
is enthalpically unstable towards decomposition to elemental silicon
and hydrogen below 95\,GPa. A decomposition to silane
SiH$_4$ together with elemental silicon and hydrogen is enthalpically
possible up to pressures of 190\,GPa. This compositional instability
could pose challenges en route to synthetization of crystalline
disilane, depending on barrier heights and on the dynamics of the
decomposition.

Yet another low-enthalpy metallic phase of disilane was found
during our MHM simulations (see Fig.~\ref{fig:Structure}a). It belongs
to the $Cmcm$ space group, and is the lowest enthalpy structure 
up to 280\,GPa.  At 200\,GPa, its conventional cell
parameters are $a=7.965$\,\AA, $b=2.705$\,\AA, and $c=4.728$\,\AA,
with one silicon atom occupying the $8e$ crystallographic site at
$(0.141,0,0)$ and three hydrogen atoms occupying $8g$, $8g$ and $8f$
sites at coordinates $(0.293,0.173,0.250)$, $(0.086,0.302,0.250)$ and
$(0,0.311,0.895)$, respectively. The hydrogen atoms are embedded
into a framework of five-fold coordinated silicon atoms. The average
silicon-silicon bond length is 2.28\,\AA, and each silicon atom is
surrounded by six hydrogen atoms at a mean distance of
1.52\,\AA.

\begin{figure}[t]            
\includegraphics[width=0.9\columnwidth]{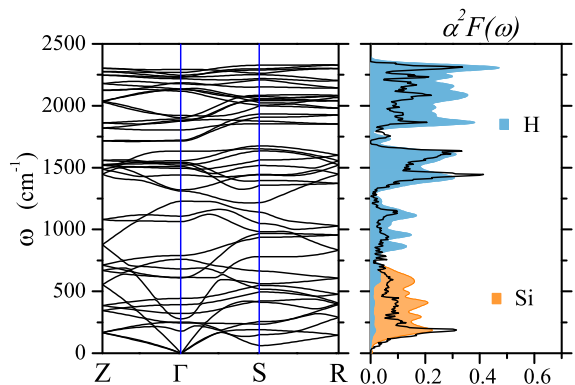} 
\caption{(Color online) 
  Left panel: Phonon band dispersion of the $Cmcm$ structure at 200~GPa.  
  Right panel: Calculated Eliashberg spectral function
  $\alpha^2F(\omega)$ (solid line) and phonon partial density of
  states (Yellow: silicon, Blue: hydrogen).}  
\label{fig:A2F_PHON}
\end{figure}

We further characterized the $Cmcm$ structure by performing
calculations of the phonon-spectrum, the electron-phonon coupling and
the superconducting transition temperature $T_c$. The phonon spectrum
and the electron-phonon matrix elements were obtained from
density-functional perturbation theory~\cite{DFPT_S.Baroni}. The
spectral function $\alpha^2F(\omega)$ was integrated over the Fermi
surface by applying the tetrahedron technique. Convergence of the above quantities was
ensured by a $16\times16\times16$ Monkhorst-Pack $k$-point sampling,
and a $4\times4\times4$ $q$-point sampling for the phonon
wave-vectors. The phonon dispersion was obtained by Fourier
interpolating the computed dynamical matrices.

The phonon band dispersion of the $Cmcm$ phase at 200\,GPa can be seen
in the left panel of Fig.~\ref{fig:A2F_PHON}, while the the partial
phonon density of states is shown in the right panel. As expected, the
low frequencies ($<$700\,cm$^{-1}$) are dominated by the vibrations of
the silicon framework whereas the high end of the spectrum extending
up to 2300~cm$^{-1}$ is solely due to the light hydrogen atoms. We
found the structure to be dynamically stable up to 220\,GPa. However,
if the pressure is increased beyond 225\,GPa a dynamical instability
arises. The phonon band dispersion at 230\,GPa can be found in the
Supplementary Material~\cite{SupplMat}. It shows an imaginary (plotted as negative)
frequency at the $\Gamma$-point, indicating an unstable phonon
mode. Following the eigendisplacements of this mode, 
which are shown in Fig.~\ref{fig:Structure}c, and then performing a full relaxation 
of the structure leads to another unreported
stable structure with $Cmc2_1$ symmetry. Compared to the $Cmcm$ phase, the silicon
framework remains essentially intact while the hydrogen atoms are
slightly displaced, partially breaking the symmetry. Due to the strong
similarities between the $Cmcm$ and the $Cmc2_1$ structures we do not
expect large differences in their phonons or superconducting properties.
A similar analysis as above has been carried out following a further imaginary frequency arising 
at the $S$-point when the pressure is increased above 260\,GPa. The resulting structure 
found by following the corresponding eigendisplacements resulted in a structure with 
$P1c1$-symmetry (see the Supplementary Materials for details on this structure)~\cite{SupplMat}.

\begin{table}
\caption{\label{tbl:McMillan} Superconducting properties of the 
  $Cmcm$ phase at different pressures. The transition temperatures were calculated using
  McMillan's formula.} 
\begin{tabular}{lcccc}
  Pressure & $\lambda$ & $\omega_\text{log}$ & $T_c$         & $T_c$ \\ 
           &           &                     & $\mu^*=0.1$  & $\mu^*=0.13$ \\
  \hline
  100 & 0.84 & 478  & 24.5 & 20.1 \\
  140 & 0.68 & 553  & 18.0 & 13.6 \\
  160 & 0.66 & 556  & 16.9 & 12.8 \\
  200 & 0.68 & 501  & 16.4 & 12.4 \\
  220 & 0.76 & 384  & 16.4 & 13.0 
  \end{tabular}
\end{table}

In order to investigate the superconducting properties of the $Cmcm$
phase, we use McMillan's approximate formula for the
superconducting transition temperature $T_c$~\cite{McMillanTC,Allen-DynesTC}.
McMillan's formula 
requires the Eliashberg spectral function $\alpha^2F(\omega)$, which
was obtained from \textit{ab initio} calculations performed with the 
\textit{abinit} code~\cite{gonze_brief_2005,gonze_abinit_2009}. In the right panel
of Fig.~\ref{fig:A2F_PHON}, the solid lines represent the Eliashberg
spectral function of the $Cmcm$ phase at 200\,GPa.  It has three main
features: (i)~low optical modes of the silicon framework, (ii)~two
intense hydrogen peaks around 1500\,cm$^{-1}$ and 1600\,cm$^{-1}$, and
(iii)~high frequency modes of hydrogen around 2000\,cm$^{-1}$.

The superconducting properties of the $Cmcm$ phase at several
pressures are summarized in Table~\ref{tbl:McMillan}, using two
different values for the Coulomb pseudopotential $\mu^*=0.1$ and
$\mu^*=0.13$. Assuming the larger of those values, the superconducting
transition temperature $T_c$ is 20.1\,K at 100\,GPa and decreases to
13.0\,K at 220\,GPa. A decreasing $T_c$ with respect to increasing
pressure is a general feature observed in hydrogen rich
materials~\cite{eremets_superconductivity_2008,PRL_PtH_Ahuja,PRB_PtH}.
We should emphasize that the $T_c$ of the $Cmcm$ phase is
approximately {\em one order of magnitude smaller} than in previously
reported structures, and that the $Cmcm$ phase is the lowest enthalpy
phase. This raises serious doubts if high-$T_c$ superconductivity will
ever be achieved in silane materials under reasonable pressure.

\begin{figure}[t]            
  \subfigure[]{\includegraphics[width=0.45\columnwidth,angle=0]{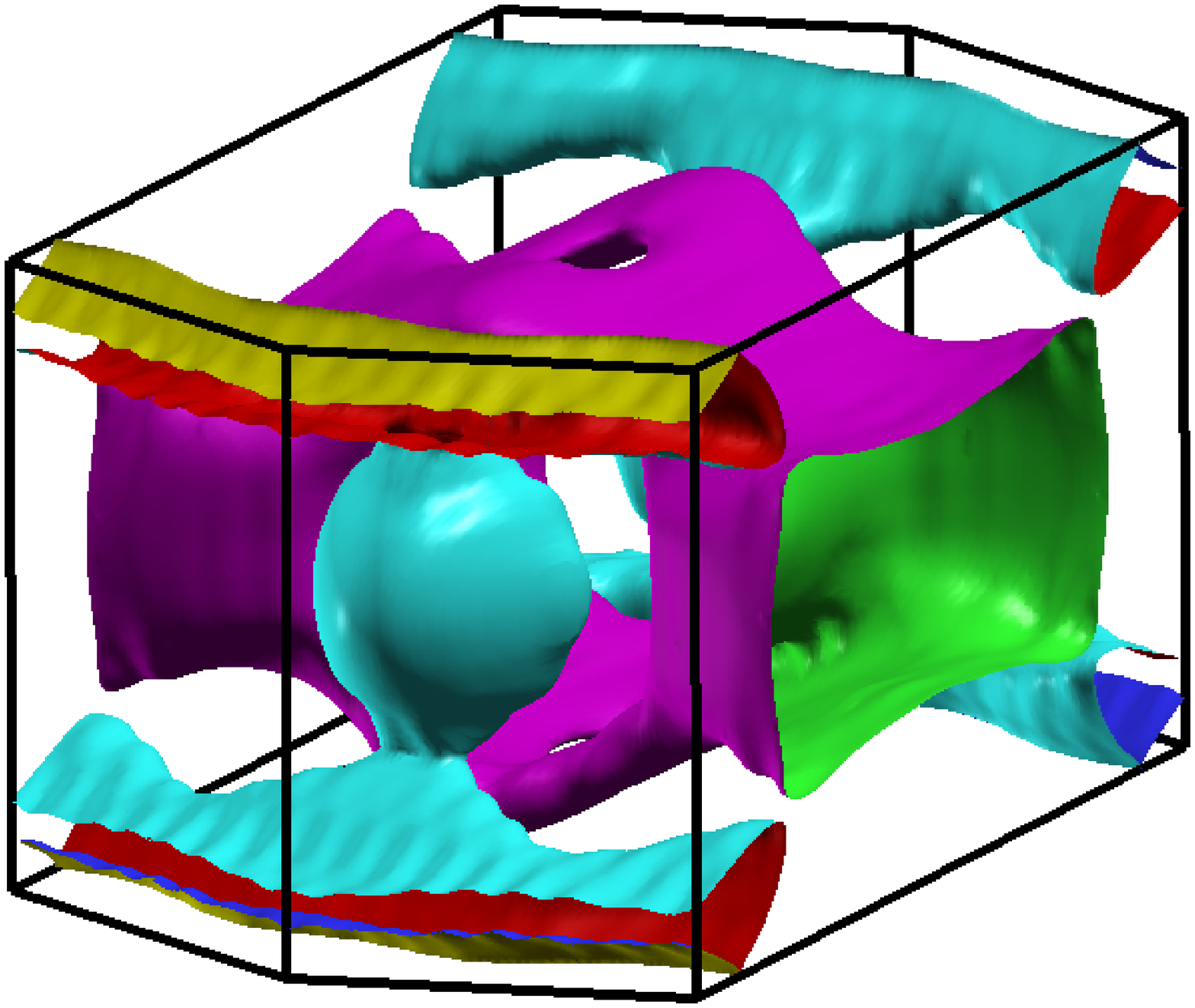}}
  \subfigure[]{\includegraphics[width=0.45\columnwidth,angle=0]{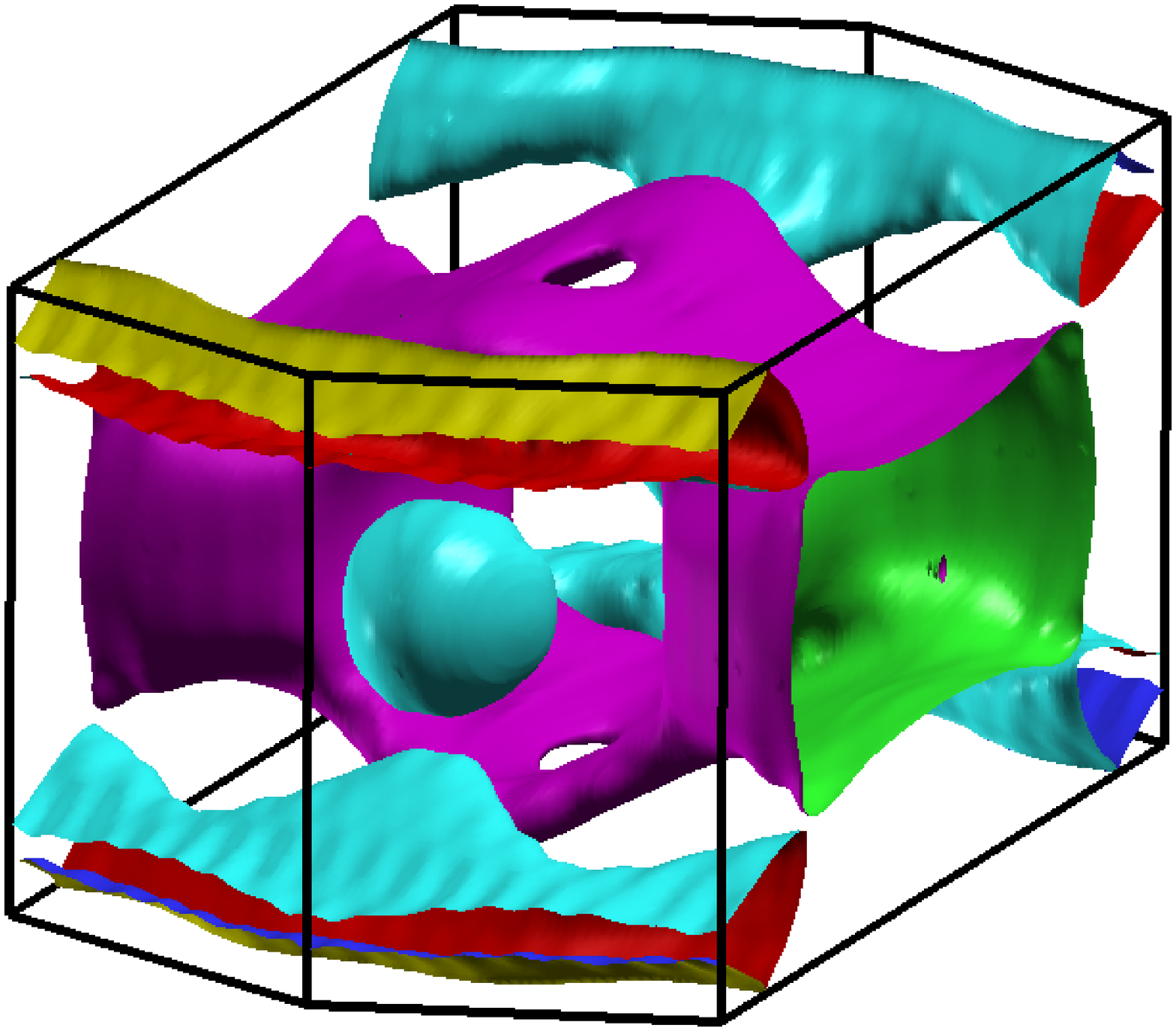}}
  \subfigure[]{\includegraphics[width=0.45\columnwidth,angle=0]{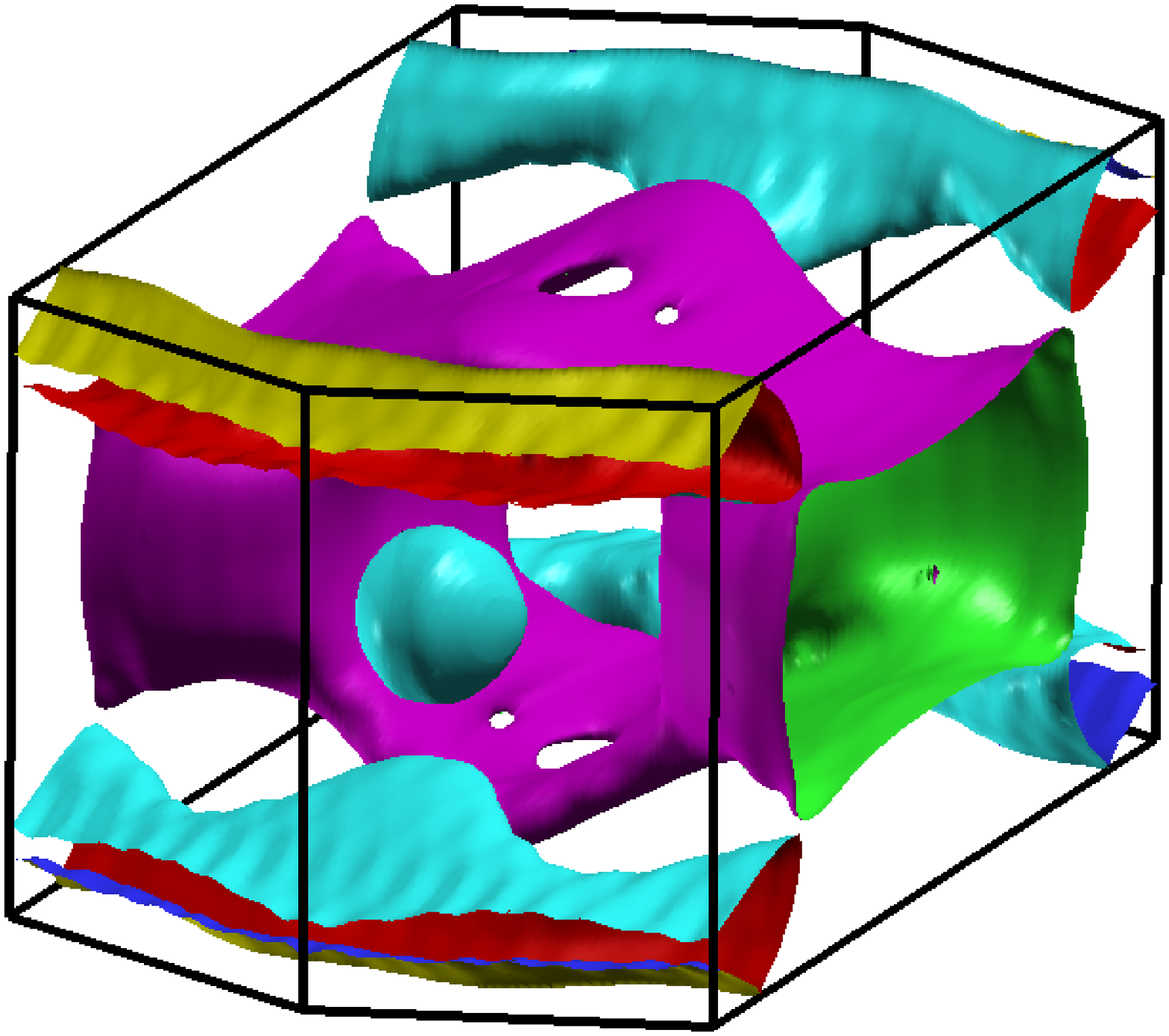}}
  \subfigure[]{\includegraphics[width=0.45\columnwidth,angle=0]{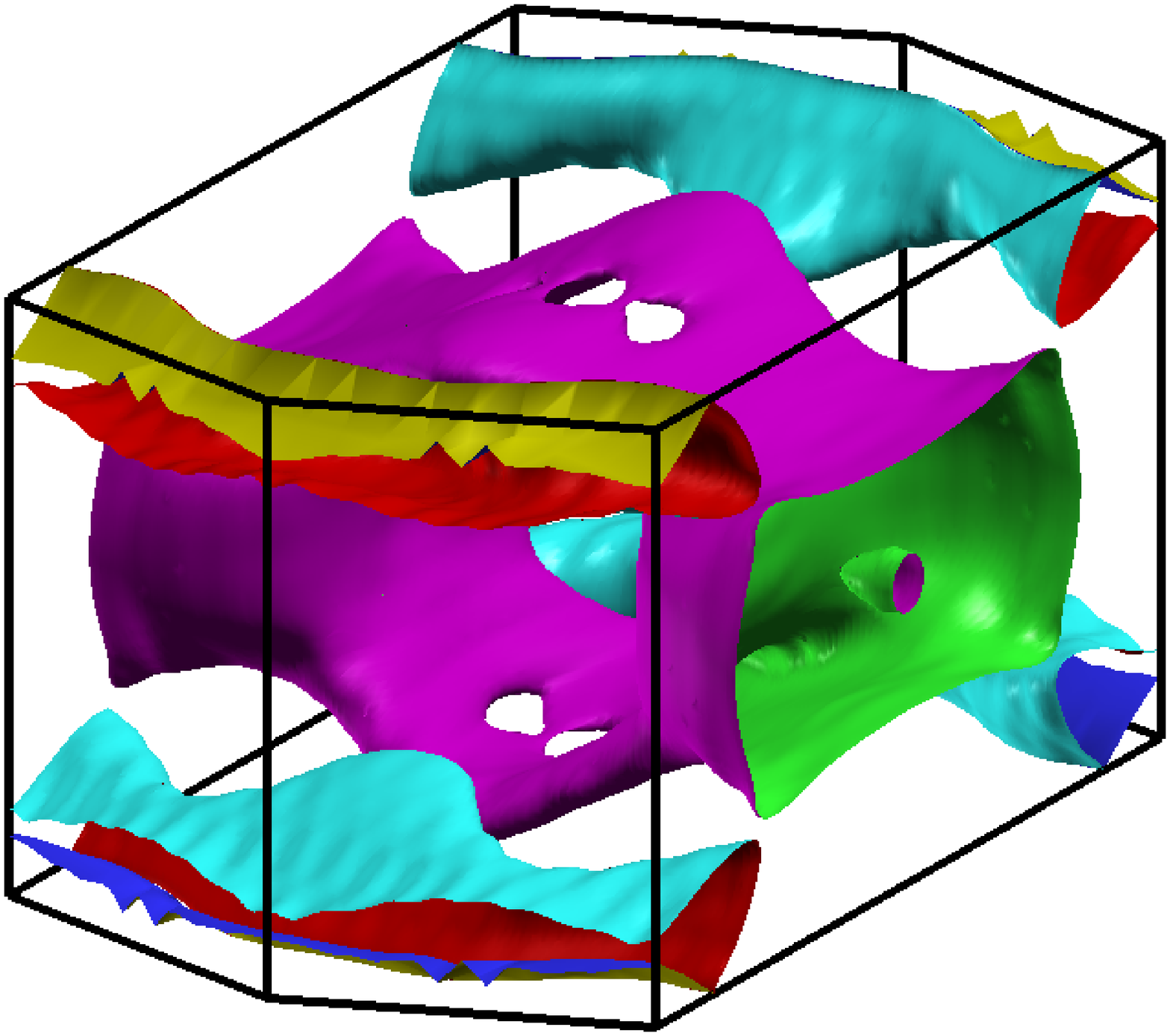}}
  \caption{(Color online) Fermi surface of the $Cmcm$ phase at several
    pressures: a) 100\,GPa, b) 140\,GPa, c) 160\,GPa, and d)
    220\,GPa.~\cite{xcrysden}} 
  \label{fig:Fermi}
\end{figure}

The superconducting properties of the $Cmcm$ phase are strongly linked
to its electronic structure. In Fig.~\ref{fig:Fermi}, the evolution of
the Fermi surface is shown as a function of pressure. Three states
cross the Fermi surface. The first (magenta) and the second (yellow)
states cover an important portion of the Brillouin zone and remain nearly 
unaltered as the pressure increases, whereas the
third (cyan) changes substantially. The contribution of this third state to the 
Fermi surface consists of spherical regions near the $\Gamma$-point. This band 
connects two main portions of the Fermi surface. Therefore, we can expected high 
superconducting values for low pressures; $\lambda=0.84$ and $\omega_\text{log}=480$ with a $T_c$ 
of 20\,K at 100~GPa. However, as the volume of the structure decreases with increasing 
pressure, this sphere-like feature of the Fermi surface is abruptly reduced. Consequently, 
at 160~GPa the superconducting parameters clearly decrease; $\lambda=0.66$, 
$\omega_\text{log}=556$ and $T_c=12.4$\,K.

In conclusion, we performed a thorough investigation of the high pressure phases of disilane
using first principles calculations. Applying our minima hopping
method to explore the potential energy surface of disilane, we found a
metallic structure which is enthalpically favorable
compared to the previously proposed structures of disilane. Additionally, the
systematic study of the superconducting properties as a function of
pressure shows that the $Cmcm$ phase possesses a moderate
electron-phonon coupling, leading to a superconducting transition
temperature in the 10--20\,K range. This result stands in sharp contrast
with the structures previously proposed of disilane under
pressure. Moreover, we observed that the transition temperature of the
$Cmcm$ structure has the tendency to decrease monotonically with
applied pressure (for $\mu^*=0.1$), which can be understood by the shrinking of a part
of the Fermi surface. This decrease of the $T_c$ is in agreement with most
theoretical and experimental results of hydrogen rich materials,
including
silane~\cite{eremets_superconductivity_2008,PRL_PtH_Ahuja,PRB_PtH}.
Certainly, this does not imply that superconductivity in hydrogen rich
materials is limited to relatively low values of $T_c$, but our results do
impose strong constraints on the possibility of high-$T_c$ superconductors 
in silicon-hydrogen systems.

Financial support provided by the Swiss National Science Foundation are gratefully acknowledged. 
SB acknowledges support from EU’s 7th Framework Programme (e-I3 contract ETSF) and MALM from the
French ANR (ANR-08-CEXC8-008-01). MA and JAFL acknowledge the computational resources provided 
by the Swiss National Supercomputing Center (CSCS) in Manno.
JAFL acknowledges the CONACYT-Mexico and computational resources provided by IDRIS-GENCI (project x2011096017) in France.

%

\end{document}